\documentclass[singlecolumn,amsmath,amssymb]{revtex4}

\usepackage{graphicx}% Include figure files
\usepackage{dcolumn}% Align table columns on decimal point
\usepackage{bm}% bold math

\begin{document}

\title{Parton Level study of high $E_T$ jets in hard QCD processes at 
LHC}

\author{M. Kaur}
\email{manjit@pu.ac.in} 
\author{Ruchi Gupta}%
\affiliation{Department
of Physics, Panjab University, Chandigarh-160 014, India.}

\begin{abstract}
Inclusive jet production will dominate the high $Q^{2}$ final states at 
the LHC.~In this work we try to estimate the up-to-date 
expectations, for high $E_{T}$ jets and their 
expected origin from the various parton-parton scattering processes.
For these studies we have used a standard Parton Distribution
Function (PDF) and simulated millions of events with the PYTHIA8 
event genertor. The results are compared with simulations for 
center-of-mass
energies of 0.9 TeV, 2.36 TeV, 7 TeV  and 14 TeV corresponding to existing and
future LHC runs.
We present some expectations for the relative
cross sections of different quark flavours which indicates that
eventually we might be able to measure the cross section for b-flavoured
jets with reasonable accuracy up to an $E_T$ of a few TeV.
\end{abstract}
\maketitle
\section{Introduction}
Hadronic collisions at high energies have been extensively studied at
various hadron colliders reaffirming that QCD gives a good description 
of strong interactions. Studies with very large momentum transfer,
inclusive hadronic final states, heavy flavours and $\gamma$-jet events
are usually considered to be an essential physics at the LHC.\\

Results from previous experiments at the Tevatron and at the
CERN  $p\bar p$ collider have demonstrated that the
'hard scattering' jet cross-section is so far well described by QCD.
Some Tevatron results of jets up to 450 GeV i.e. 
$X_{T}(=E_{T}/E_{beam})\approx$ 0.5\cite{mangano1} show
that the agreement between theory and data is at the level of 30\%
for cross-sections between $E_{T}$ = 20 GeV to $E_{T}$ = 450 GeV.
Some dependence on the chosen set of Parton Distribution Functions
(PDF) is also shown in these papers. The indications of some excess at
large  $X_{T}$ are currently attributed to the uncertainty on the gluon 
density
at $X_{T}$.\\

Within QCD, the only fundamental quantity that needs to be known
is $\alpha_s$, the QCD coupling constant. So far there are  no
concrete ideas on how to improve the knowledge
of $\alpha_s$ at the LHC. Nevertheless, it is obvious that the
LHC experiments will allow accurate consistency checks with
QCD predictions for the $E_{T}$ dependent jet cross sections.
 $E_{T}$ values reached at Tevatron (450 GeV) can be increased by
roughly a factor of ten well into the multi TeV domain. The predictions 
for these dominant QCD
jet cross-sections are also important as
potential backgrounds in the search for new phenomena.
Even though the expectations summarized in this note are based on 
leading order
QCD calculations
only, we expect that they provide a useful guidance for future NLO calculations
and more detailed studies about jet physics at the LHC.\\

The structure of the proton at short distances is
crucial to predict jet cross-sections and variations from the different 
types of partons (quarks and gluons). The
contributions from the different parton--parton
collisions are investigated as a funtion of $E_{T}$ and jet rapidity.
The complex dynamics
of processes involving hadrons in the initial state needs to be
investigated in detail in order to measure and interpret the physics at 
the
highest possible jet $E_{T}$, which corresponds to distance scales of
$O(10^{-17}$cm) much smaller than for any other LHC process.
In order to study variations as a function of the center of mass energy,
some distributions are shown as a function of the scaling variable 
$X_{T}$.
Finally we give also cross sections of quark final states in the various
flavours. The results obtained  show that a measurement of the 
b-flavoured jets
should be possible up to an $E_{T}$ of a few TeV.\\

The results presented in the following need to be
improved using Next-to-Leading-Order calculations.Some generators such 
as HERWIG \cite{Corcella} with MC$@$NLO \cite{Frixione} and 
POWHEG \cite{Nason} have included the NLO calculations. However, the 
detailed subprocess calculations in different $E_{T}$ intervals are still not 
posiible.
Furthermore possible systematics from jet fragmentation and the
actual reconstruction of jets from
the observable mesons and baryons, from hadronization schemes, from
diffractive processes and the underlying event structure need also to be 
investigated during the coming years. This requires
the precise information of parton densities, distribution functions and
related uncertainties, particularly when the initial states are composed
of hadrons. The use of high $E_T$ data which are sensitive to a variety 
of exotic phenomena will be limited by the presence of all these 
potential uncertainties.
\section{Jet Production Rates}
Jet production dominates by far all hard processes (high $E_{T}$) in
hadronic collisions. The rapidity and the $E_{T}$ distributions also
depend on the parton distribution function (PDF), $\it f_{q,g}(x, Q^2)$.
Especially the relative contributions from the gluon--gluon, 
quark--gluon
and quark--(anti)quark scattering vary strongly as a
function of both $E_{T}$ and rapidity. It is thus interesting to study
the independent contributions for the different partonic components of 
the proton.
We compute the relative contributions of different initial state partons 
to the
hard QCD jet
cross-sections. Table 1 shows the fraction of qq, qg and gg processes 
for $X_{T} > 0.5$ using
14 million events for 14 TeV, 7 million events each for 7 TeV,
4.6 million events for 2.36 TeV and 1.8 Million events for 0.9 TeV 
energies ( 100K events for each $E_{T}$ bin) corresponding to various 
LHC existing and proposed runs, using
PYTHIA8 \cite{pyth} and CTEQ6L as Parton Distribution Function set.
For the purpose of the study presented here the particular choice of the 
PDF
is not really relevant but needs to be investigated in the future.

%\par
\begin{table}[htb]
\caption{Fraction of Quark/Gluon contributions to the jet
cross-section with a fractional transverse energy $X_{T} > 0.5$ }
\label{tab:page_layout}
\begin{center}
 \large
%\vspace*{0.2cm}
\begin{tabular}{|c|c|c|c|c|}
\hline
  Energy (TeV)  &   qq   &     qg & gg   \\
\hline
14   &  $\sim 81.21\%$  &  $\sim 17.92 \%$ &  $\sim 0.86 \%$  \\
7    &  $\sim 80.29\%$  &  $\sim 18.80 \%$ &  $\sim 0.90 \%$  \\
2.36 &  $\sim 79.65\%$  &  $\sim 19.40 \%$ &  $\sim 0.95 \%$  \\
0.9  &  $\sim 77.2 \%$  &  $\sim 21.16 \%$ &  $\sim 1.17 \%$  \\
\hline
\end{tabular}
\end{center}
\end{table}
%\vspace*{0.1cm}
\normalsize
At $X_{T} > 0.5$, one finds that 77--81\%
of the jets are produced by collisions involving only
initial state quarks. The remaining jets, between 18--21\%, are from 
quark--gluon collisions
while only about 1\% of these high $E_T$ jets come from gluon--gluon 
scattering.
A comparison for the different center-of-mass energies shows
that the fraction of quark--quark collisions increases with center of 
mass energy.
Previous results from jet studies at Tevatron with $E_T$ = 20 - 450
GeV, show that about 20 \% uncertainty on high $E_T$ jet rates comes
from gluon-induced processes. Our simulations for the LHC indicate that 
the
contributions of quark--gluon and gluon--gluon processes at very high 
$X_{T}$ are
slightly smaller than at Tevatron energies.
\section{Inculsive jet cross-sections}
It is anticipated that in the high luminosity period, LHC will produce 14 TeV 
p-p collisions with a luminosity $\sim 10^{34}
cm^{-2}$ $s^{-1}$, corresponding to an integrated luminosity of up to 
100 fb$^{-1}$ per
year. This will allow to reach accurate statistical precision well into 
the multi TeV range.\\
%\par
\begin{figure}[htb]
  \begin{center}
{\includegraphics* [scale=0.40] {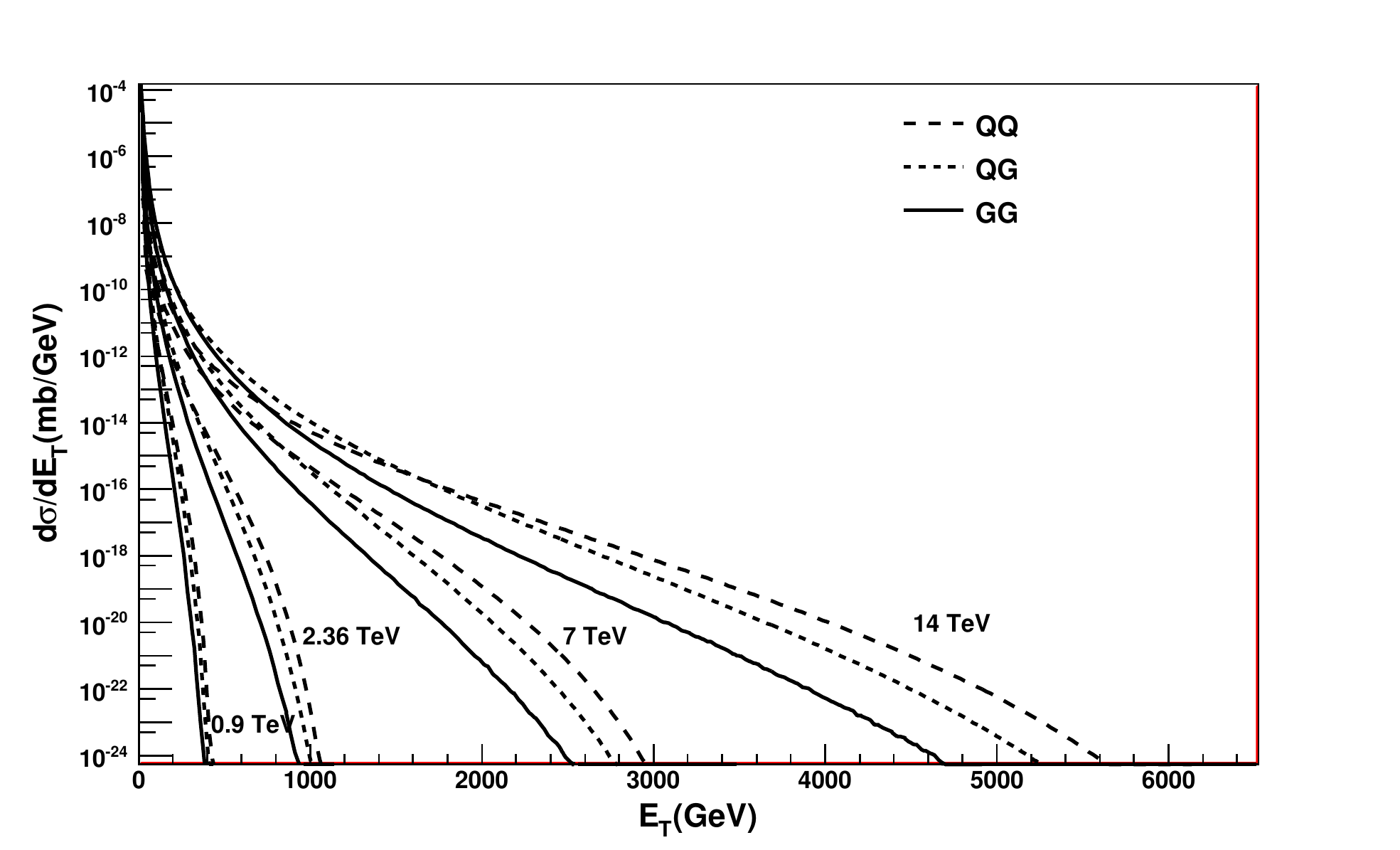},
\includegraphics* [scale=0.40] {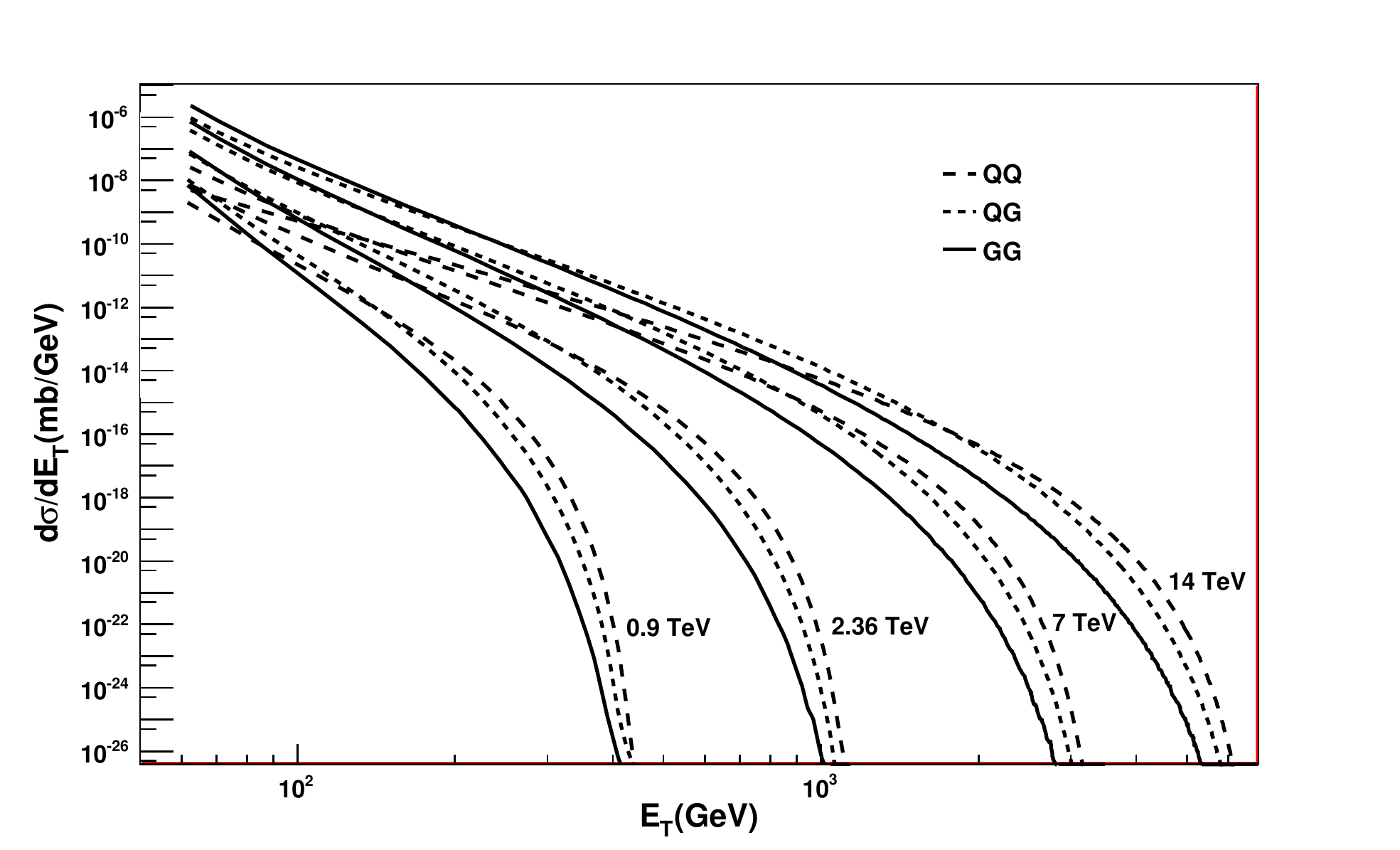}}
\caption{Inclusive jet cross-section of qq, qg, gg scattering at
 different c.m. energies.}
\label{ fig:1}
  \end{center}
  \end{figure}

Figures 1a and b show the total jet cross section as a function of 
$E_{T}$, with a linear
(a) and log scale (b) for $E_{T}$.
The different types of parton collisions at 14 TeV and for 
comparison the
corresponding curves at 7 TeV, 2.36 TeV and 0.9 TeV are also shown.
In order to make the expected approximate scaling more evident, it is 
better to
study the cross section as a funtion of $X_{T}$.
Furthermore, it is interesting to see the relative contributions to the 
jet cross section for the three partonic collisions. In Fig.2, these fractions 
are shown as a function of $X_{T}$, demonstrating the approximate scaling 
for the different center of mass energies. For clarity the same fractions 
are also shown separately and as a function of $E_{T}$ in Figures 3(a-c).  
\begin{figure}[htbp]
  \begin{center}
{\includegraphics* [scale=0.45] {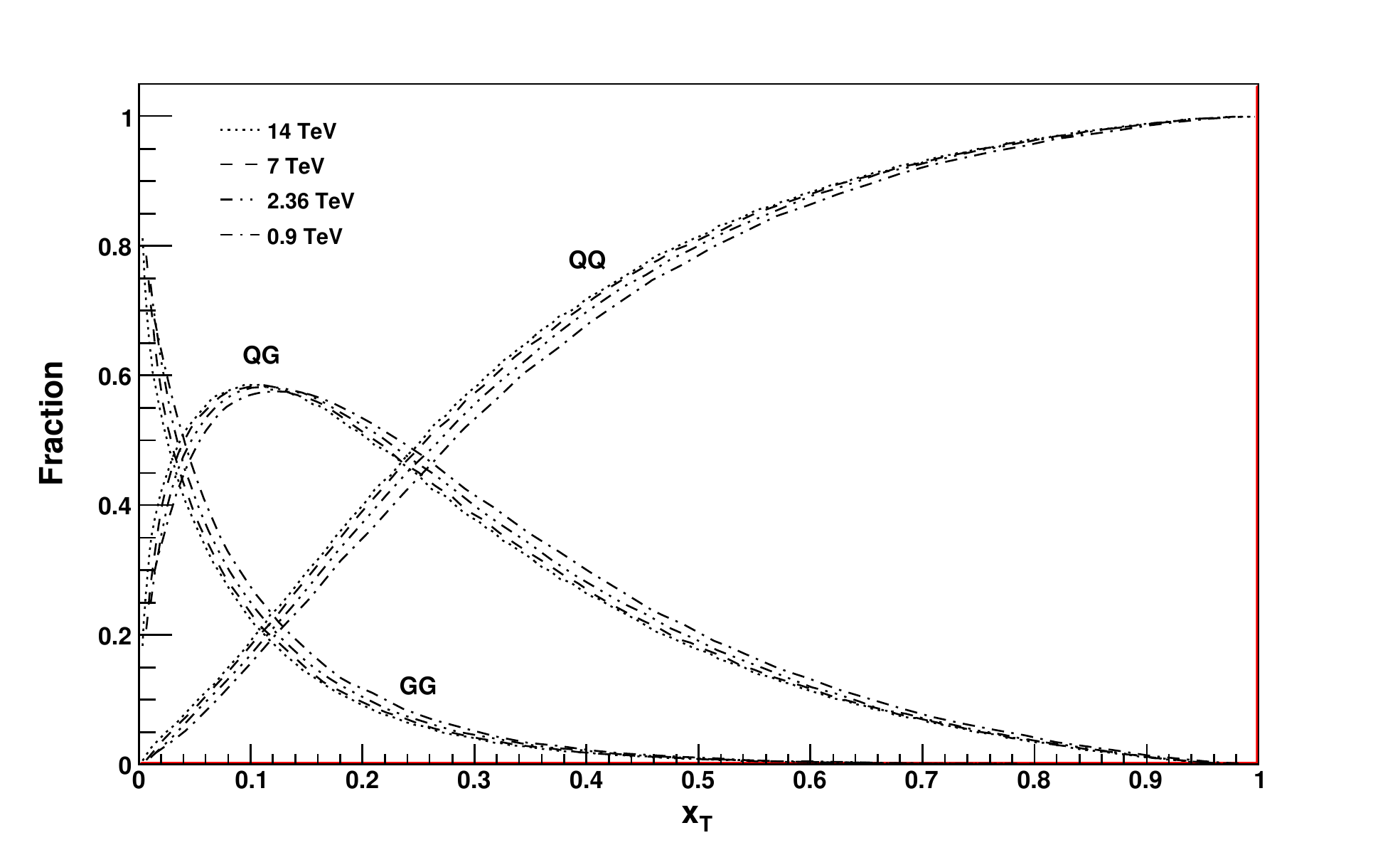}}
 \caption{Contributions to jet cross-sections from different quark 
and gluon scatterings.}
\label{fig:2}
  \end{center}
\end{figure}

%\par
\begin{figure}[htbp]
  \begin{center}
{\includegraphics* [scale=0.40] {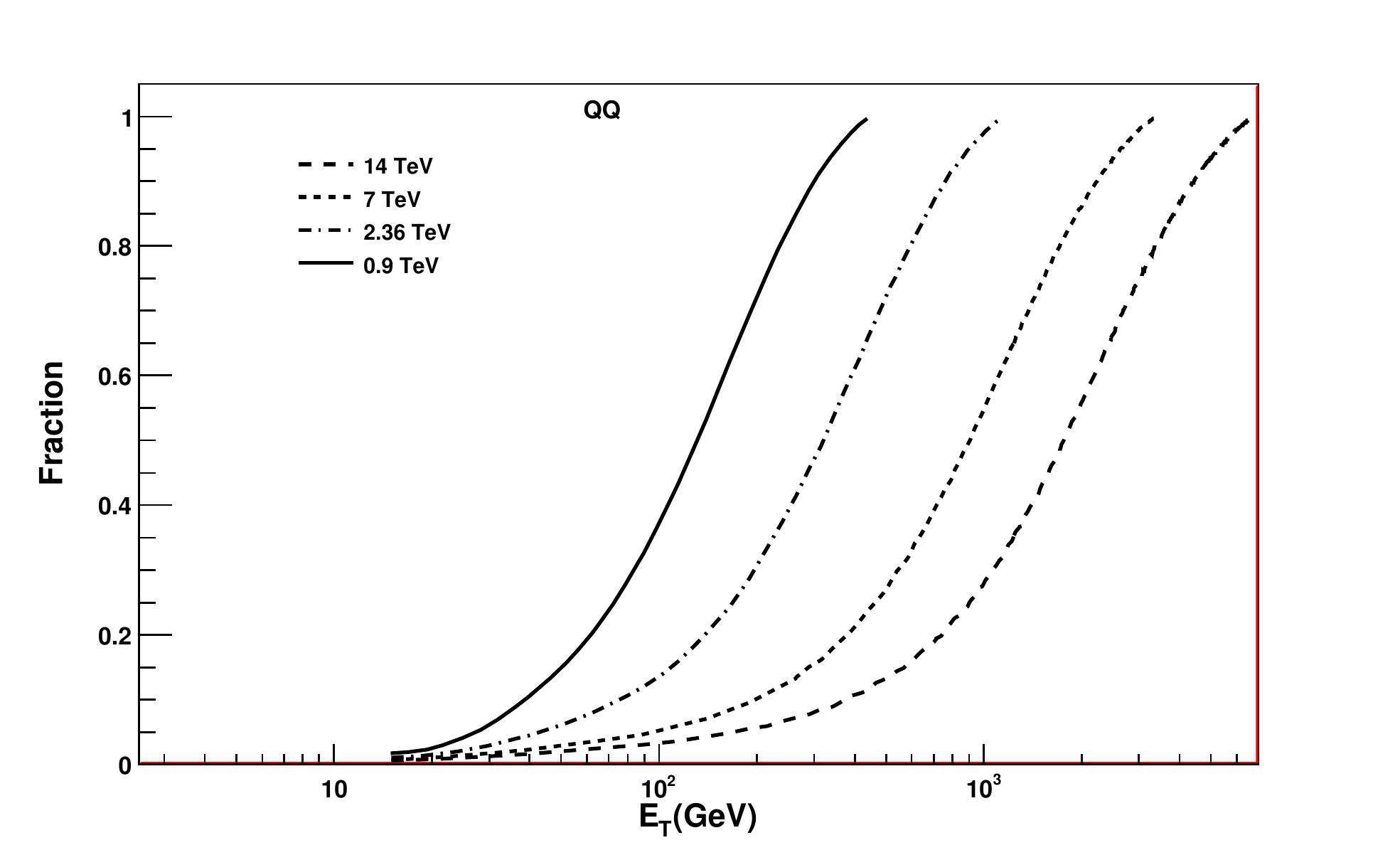}}
{\includegraphics* [scale=0.40] {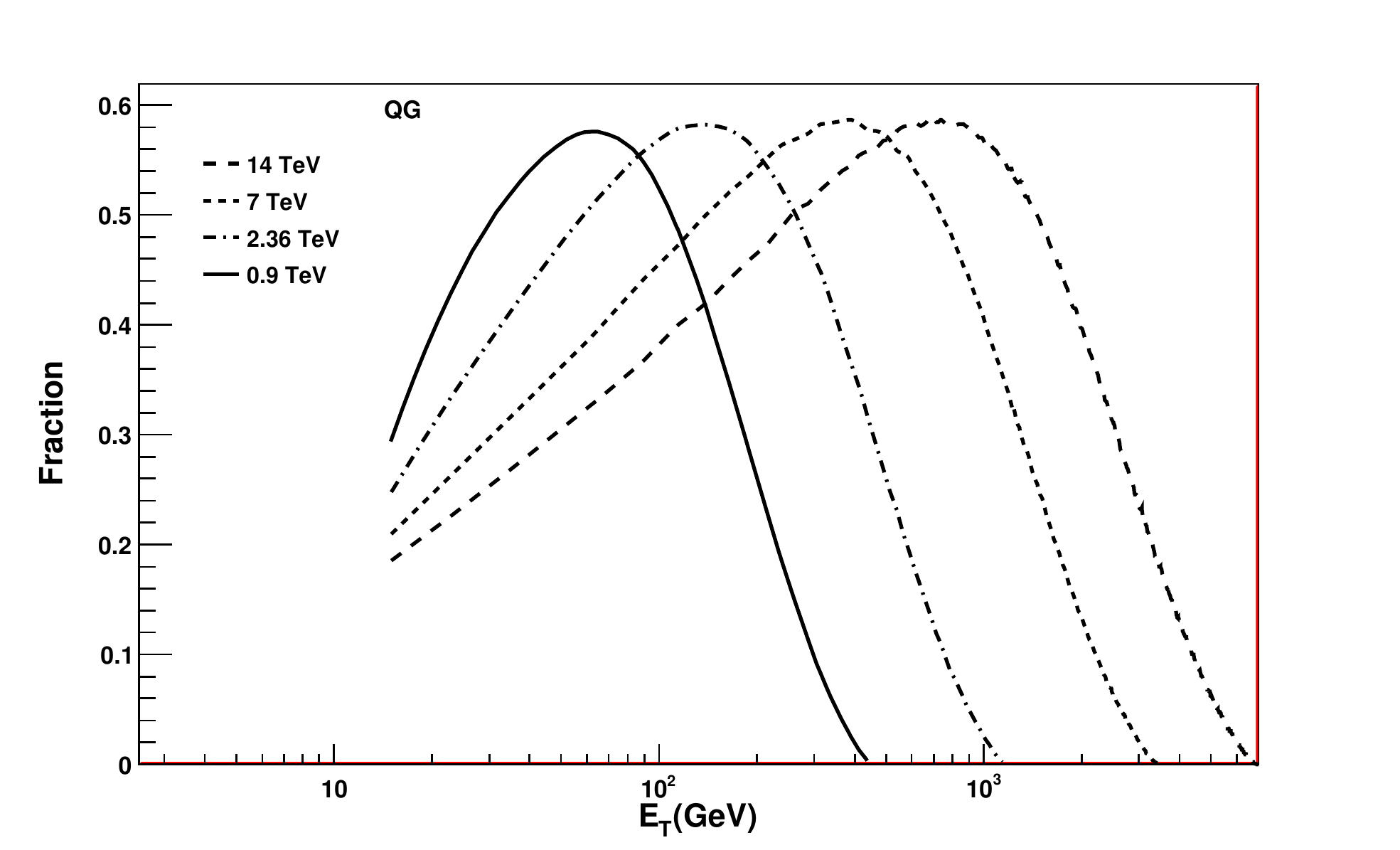}}
{\includegraphics* [scale=0.40] {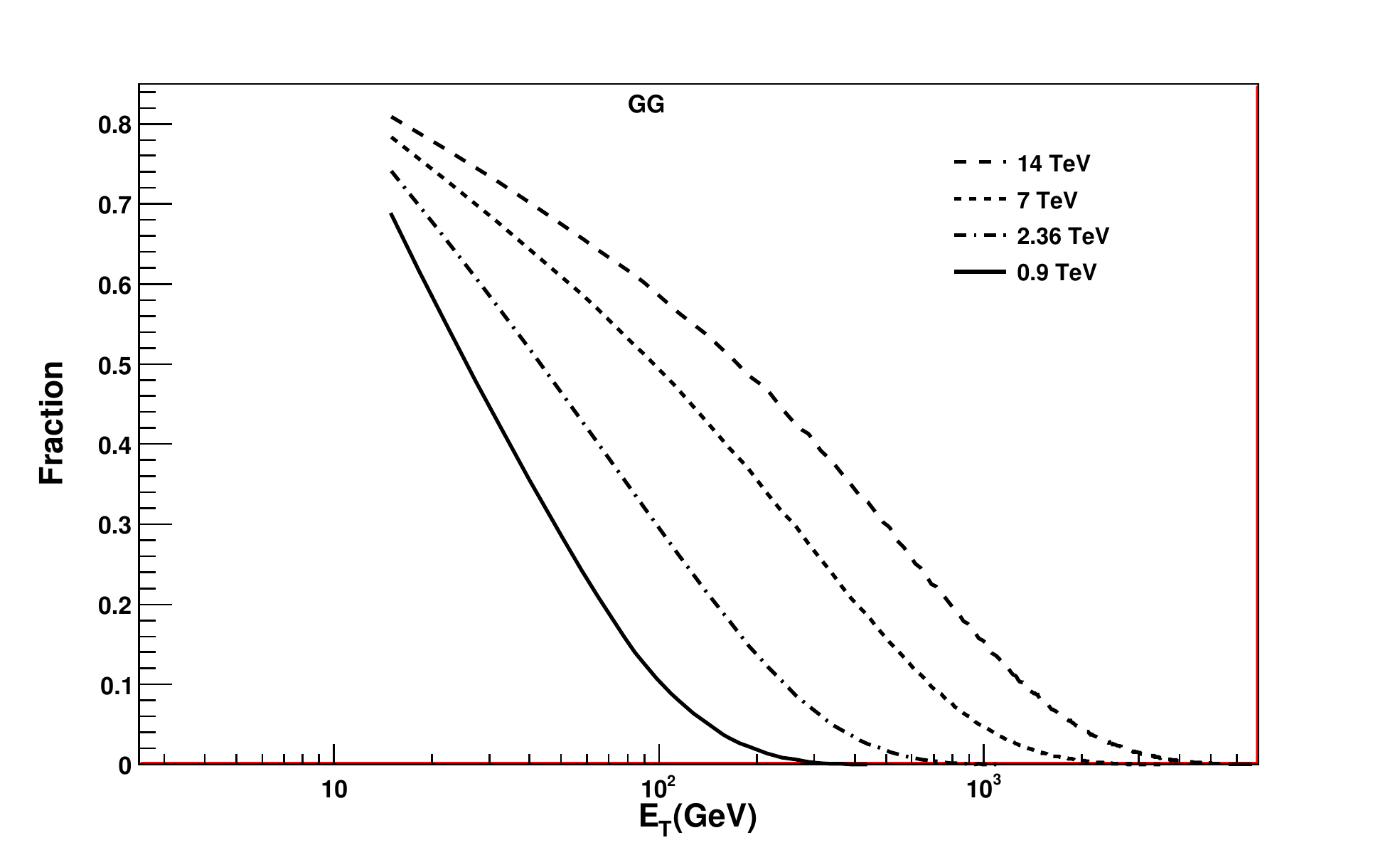}}
 \caption{Jet cross-section fractions from qq, qg, gg
scattering at different c.m. energies}
\label{fig:3}
  \end{center}
\end{figure}

At LHC, ATLAS and CMS detectors aim to achieve a high jet-energy resolution. It 
is also
envisaged that the cross section measurements at very high $E_{T}$ jets 
have in principle
the sensitivity to observe new physics like ``quark compositeness'' up to 30--40 TeV  
\cite{atlas}, assuming that the predictions from QCD calculations can be 
done with sufficient
accuracy.This needs to be demonstrated with detailed studies.\\

\begin{figure}[htbp]
  \begin{center}
{\includegraphics* [scale=0.40] {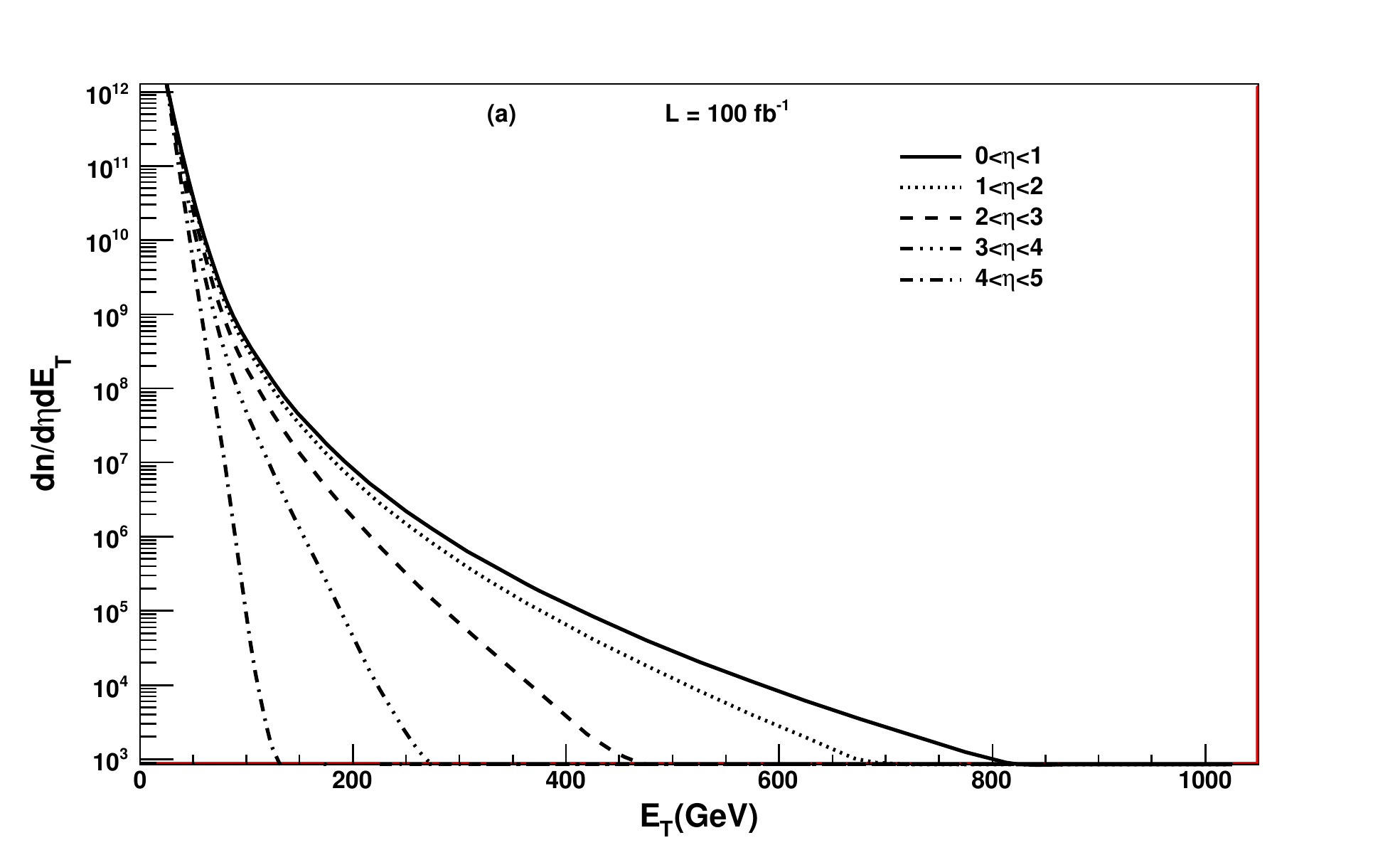}}
{\includegraphics* [scale=0.40] {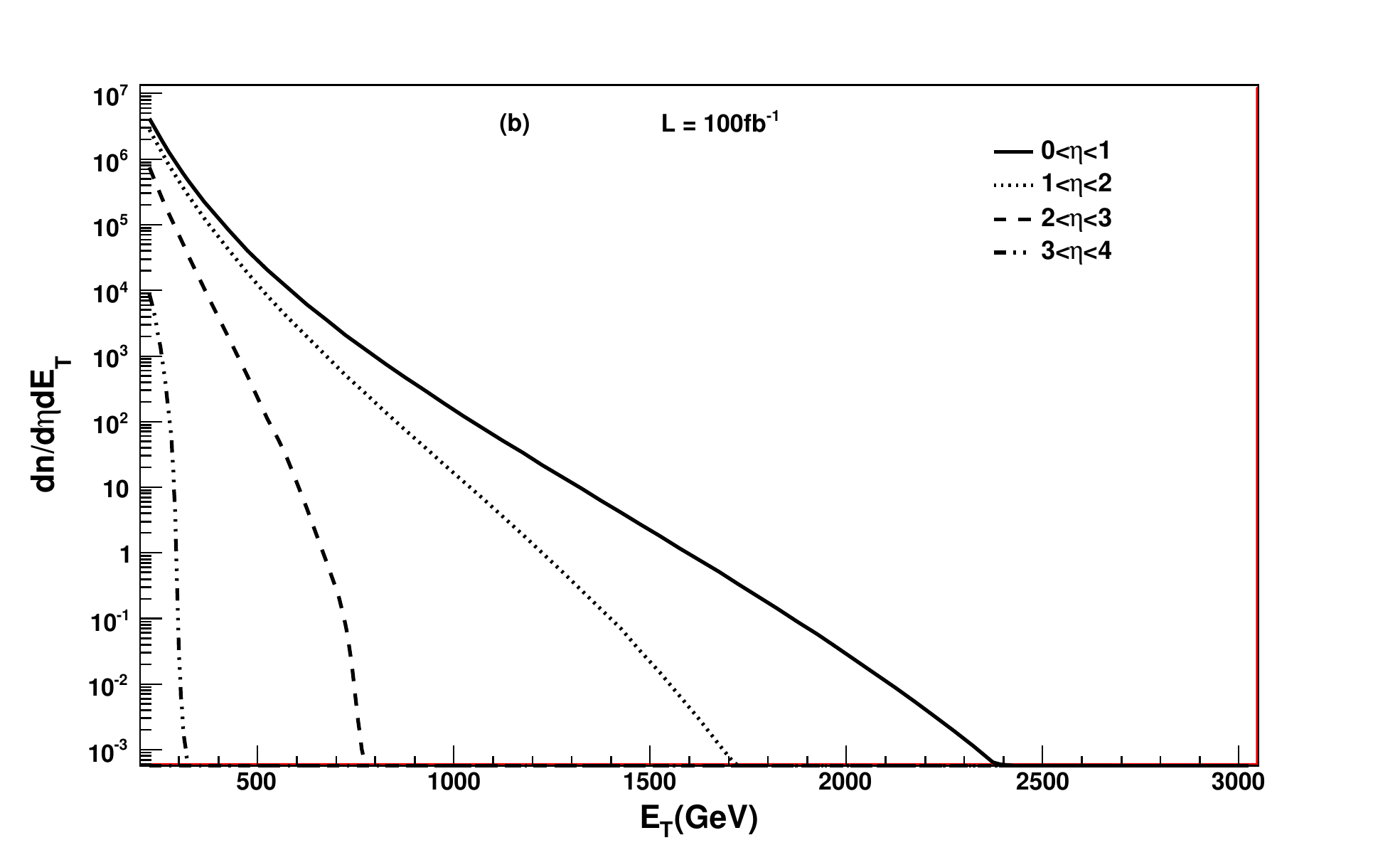}}
{\includegraphics* [scale=0.40] {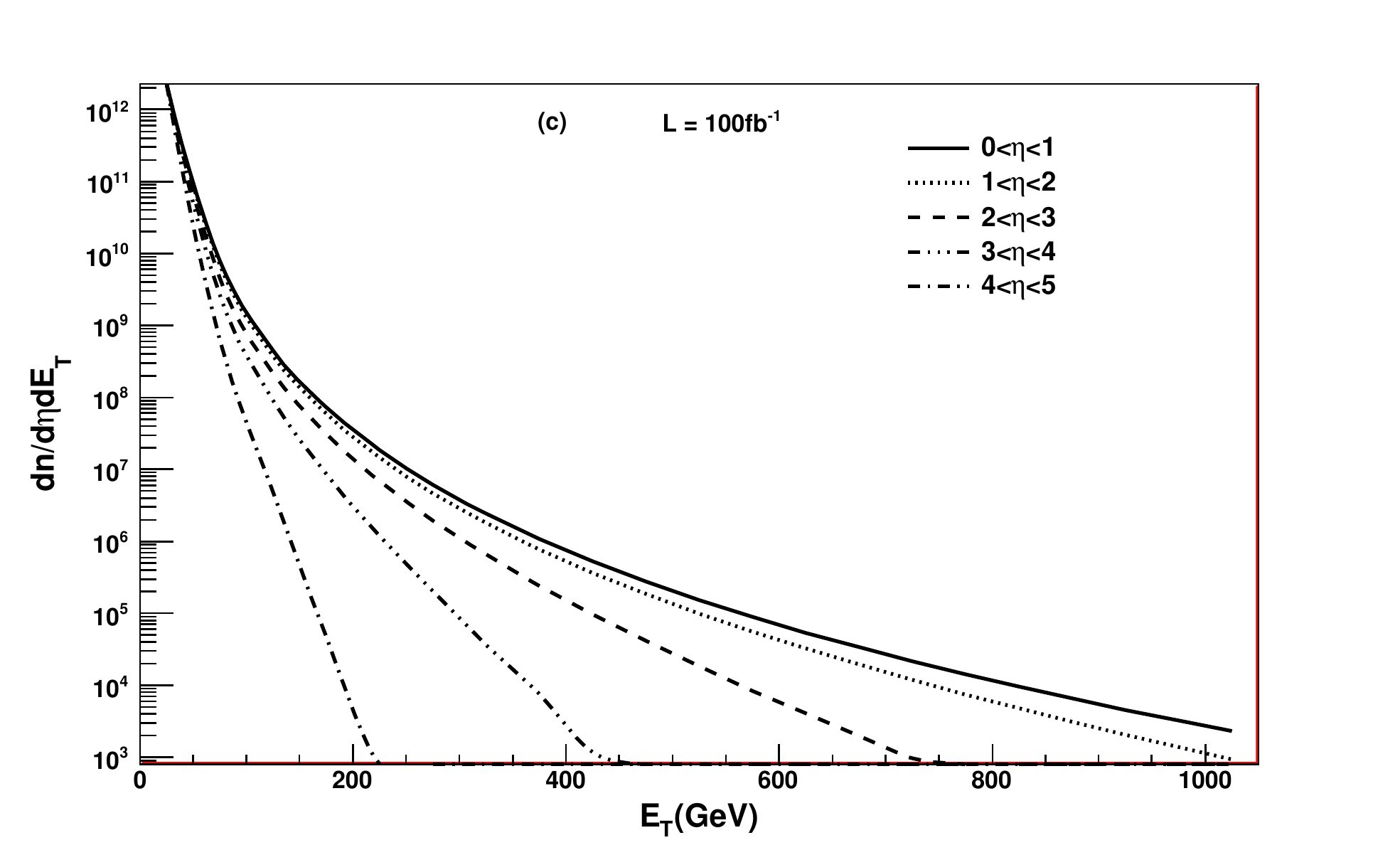}}
{\includegraphics* [scale=0.40] {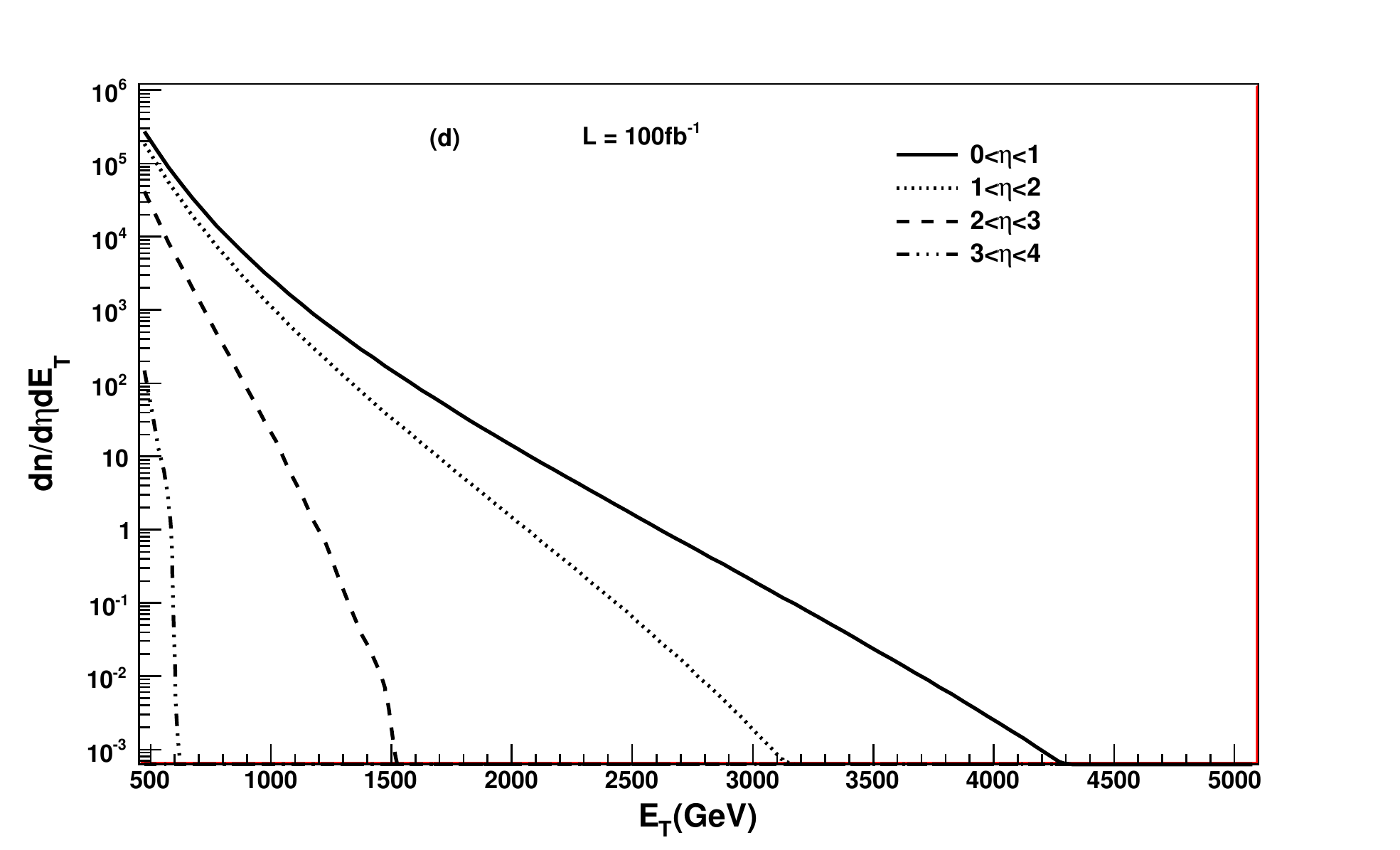}}
 \caption{Jet cross-sections for different rapidity intervals}
\label{ fig:4}
  \end{center}
\end{figure}

Figures 4(a-b) and figures 4(c-d) show the expected number of events 
for L = 100 fb$^{-1}$ at 7 TeV and at 14 TeV c.m.energies, but now
separated into different rapidity regions. 
Both figures demonstrate that the super
high $E_{T}$ jets will be produced essentially only in the small 
rapidity region, corresponding to the central part of the detector.\\

Experiments at LEP and Tevatron
\cite{Abe}\cite{Abbott}\cite{Holck} have studied also the inclusive 
charm
and bottom quark cross section by selecting samples using the known jet 
tagging
methods.
With the expected b-tagging projected capability of the tracking detectors
at LHC and the possibility to measure inclusive muon production from charm and 
beauty decays,
it might eventually be possible to measure these flavour dependent cross 
sections.

\begin{figure}[htbp]
 \begin{center}
{\includegraphics* [scale=0.40] {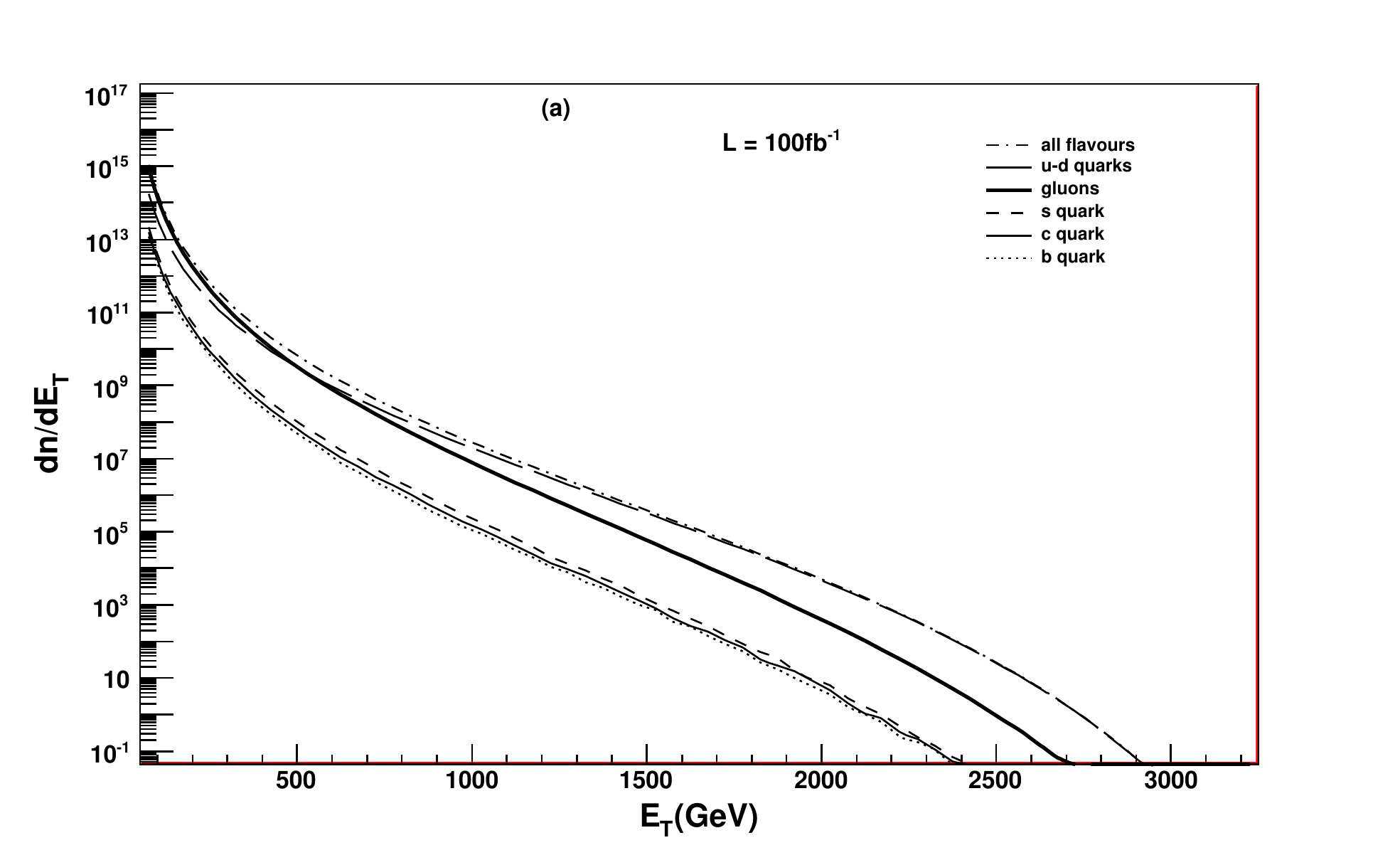}}
{\includegraphics* [scale=0.40] {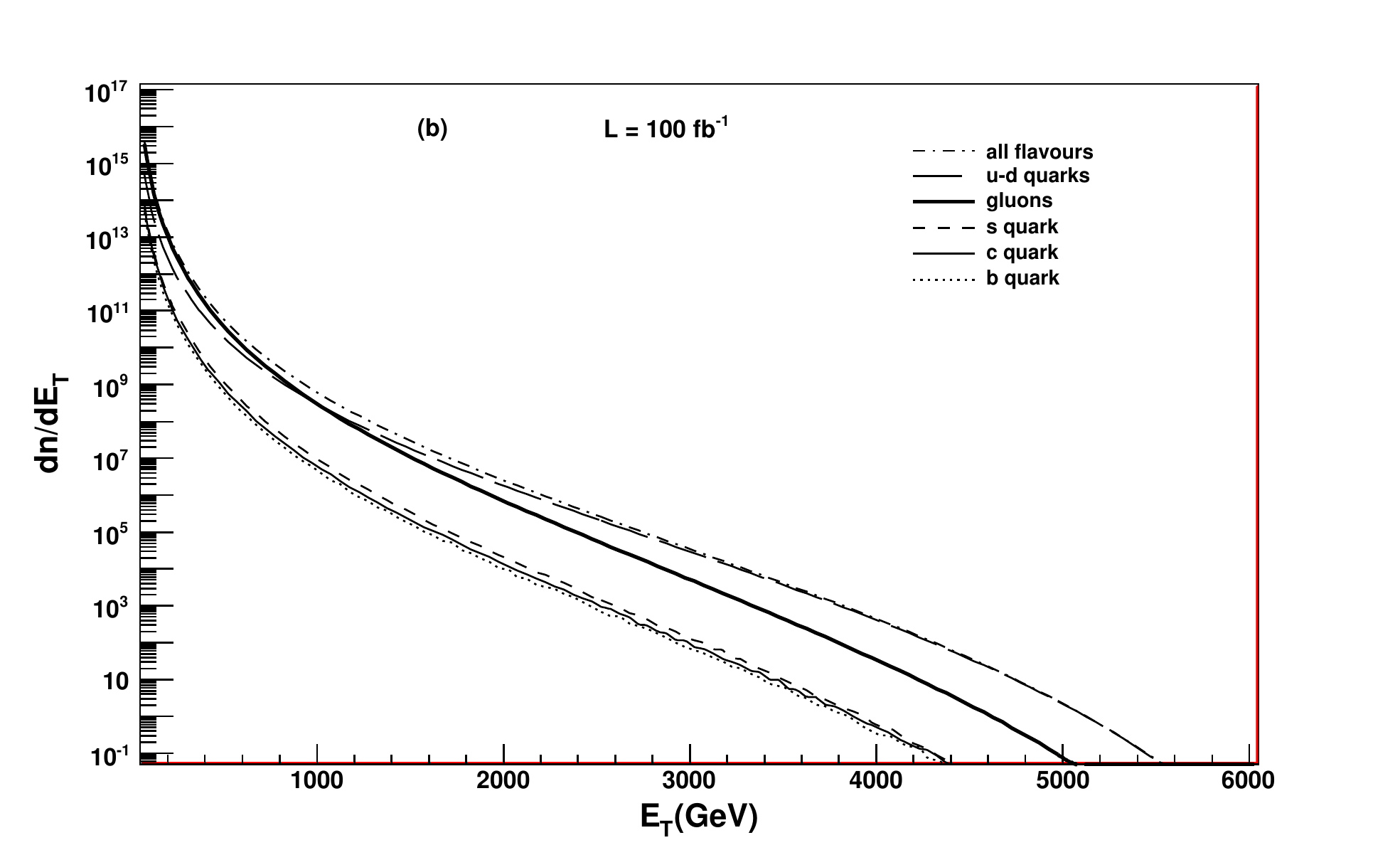}}
\caption{LHC jet cross-sections for different jet flavours}
\label{ fig:5}
  \end{center}
\end{figure}

%\section{Outlook}
\begin{figure}[htbp]
  \begin{center}
{\includegraphics* [scale=0.40] {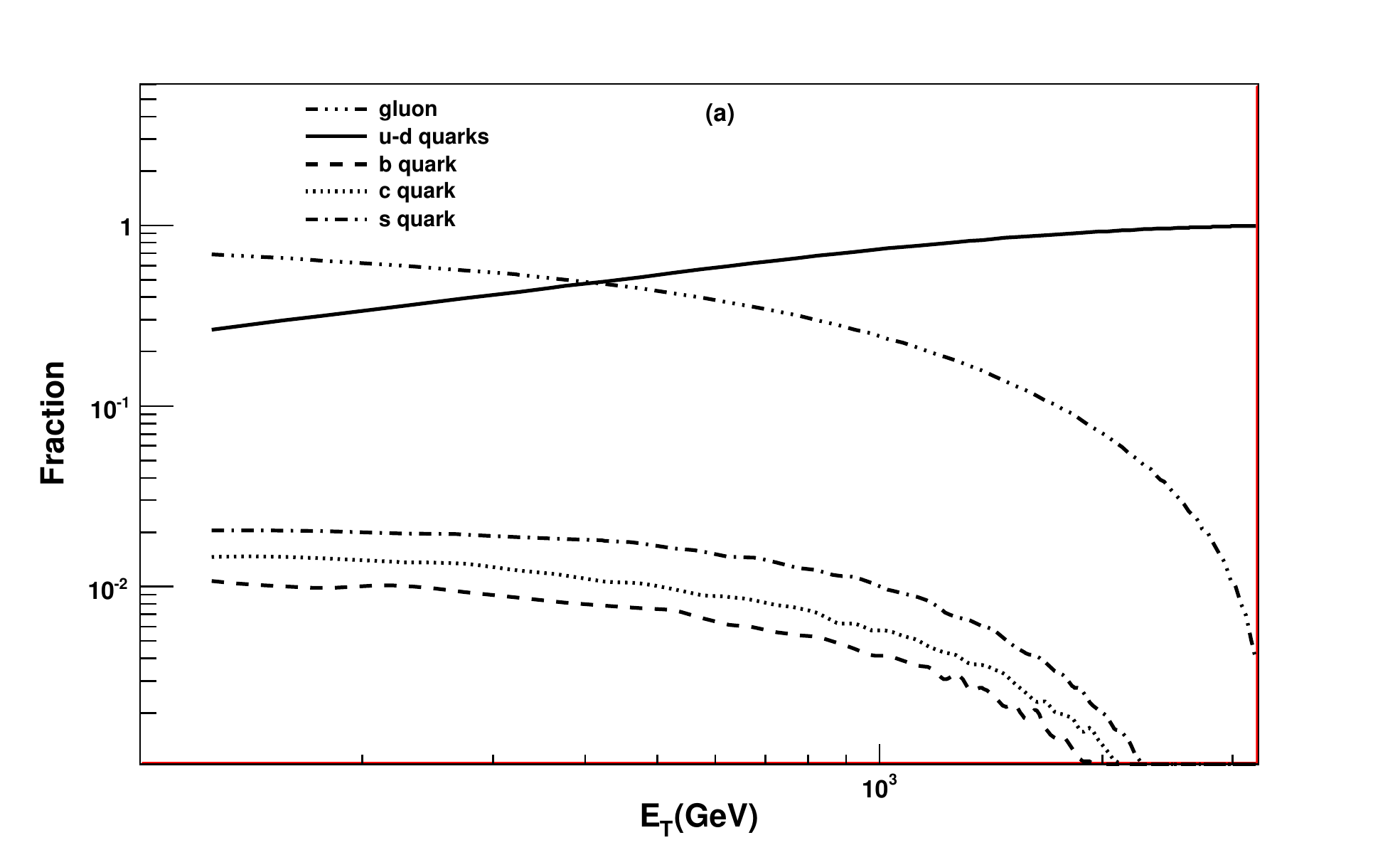}}
{\includegraphics* [scale=0.40] {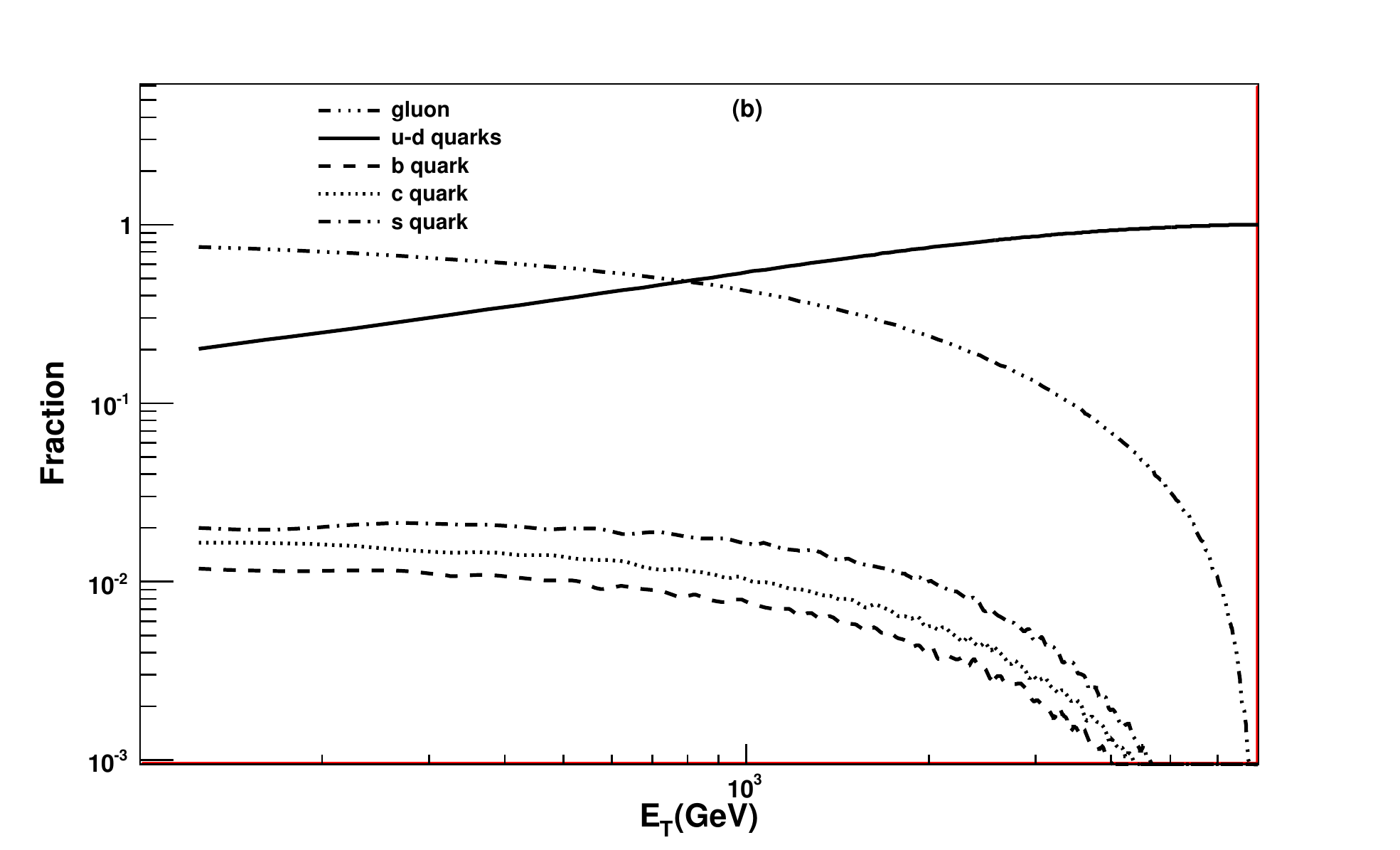}}
 \caption{Cross-section fractions for different jet flavours}
\label{ fig:6}
  \end{center}
\end{figure}

\section{Conclusion and Outlook}
 We thus present this parton level jet studies with cross sections and
fractions for the different quark flavours as a function of jet $E_{T}$ 
as shown in figures 5(a-b) and figures 6(a-b) which show that above an $E_{T}$
 of a few hundred GeV, about 2\% of all jets are beauty flavoured
and about 3\% are charm flavoured. For a jet $E_T$ larger than a few
hundred GeV, these fractions appear to be essentially $E_{T}$ 
independent up to about 1-2 TeV.
Thus, combining the large cross sections and the expected b--tagging 
capabilities of the detectors,
optimistically one might perhaps reach efficiencies of 50\% for
b--jets and 1\% for backgrounds.
Thus, the potential b--signal to background ratio should not be too 
different from 1:1.
It might thus be interesting in the near future to investigate this 
possibility in a complete ATLAS and CMS simulation.\\

The results presented here are intended to give a qualitative
up to date picture of high $E_{T}$ jets and  cross sections at the LHC.
It is obvious that these results can only be considered as a beginning
of QCD studies. The next steps should include
the effects of higher order QCD corrections and various detailed 
experimental
studies, for which one needs to investigate especially systematics
from jet reconstruction and jet resolution, including jet 
fragmentation effects,
jet algorithms and uncertainties from PDF's. HERAII data may provide improved 
statistical and systematical precision \cite{Aaron}.
It seems obvious that this currently almost ``not so well known''
area needs to be understood in detail.

%\newpage

%------------------------------------------------------------------------------
\pagebreak

% \end{flushleft} }

\end{document}